\documentclass[aps,prl,reprint,superscriptaddress,floatfix]{revtex4-2}

\usepackage{amsmath,amssymb}
\usepackage{graphicx}
\usepackage{tikz}
\usetikzlibrary{arrows.meta,positioning,calc}
\usepackage{hyperref}
\usepackage{mathrsfs}
\usepackage{slashed}
\usepackage[utf8]{inputenc}
\usepackage[normalem]{ulem}
\begin{document}

\title{A Common Type II Origin of Galilean and Carroll Superstrings}

\author{Mehmet Ozkan}
\email{ozkanmehm@itu.edu.tr}
\affiliation{Department of Physics,
Istanbul Technical University,
Maslak 34469 Istanbul,
T\"urkiye}

\author{Utku Zorba}
\email{utku.zorba@iuc.edu.tr}
\affiliation{Department of Engineering Sciences,
Istanbul University-Cerrahpa\c{s}a,
Avc\i lar 34320 Istanbul,
T\"urkiye}

\begin{abstract}
We show that the relativistic Type-II
superalgebras, once the fundamental-string winding charge is retained
alongside the momentum, admit a single non-Lorentzian contraction
with two invariant translation ideals, whose complementary quotients yield
Galilean and magnetic-Carroll superalgebras. The complementary vector charge required for this structure is thus intrinsic to string theory. The IIA
realization is particle-like and the IIB realization string-like, a
distinction fixed by chirality, and spacelike T-duality relates the
two while preserving each non-Lorentzian structure. The same
contraction is finite on the doubled Green--Schwarz action, including
all fermion terms and local kappa symmetry, where the two
supersymmetries appear as the two polarizations of a single worldsheet
theory. These results unify two distinct
non-Lorentzian corners of string theory within a common Type-II
origin and open a route toward a broader web of non-Lorentzian
dualities and possible holographic generalizations.
\end{abstract}

\maketitle
\allowdisplaybreaks

The celebrated Bronstein cube provides a clear illustration of how quantum gravity can be approached along different routes in the space spanned by the three fundamental constants of nature~\cite{Okun1991}.
Along the axis of the speed of light, string theory admits limits in which Lorentz symmetry is replaced by Galilean or Carrollian symmetry~\cite{LevyLeblond1965,BacryLevyLeblond1968}.
The resulting theories are self-contained, with their own spectra and their own dynamics.
Of particular relevance here is nonrelativistic string theory, obtained by fine-tuning the Kalb--Ramond field to its critical value while sending the speed of light to infinity~\cite{GomisOoguri2001,Danielsson:2000gi}.
This theory is unitary, has a well-defined perturbative expansion~\cite{GomisOoguri2001}, and supports a rich set of dualities, including T-duality~\cite{Bergshoeff:2018yvt} and branched $SL(2,\mathbb{Z})$ duality in type-IIB string theory~\cite{Bergshoeff:2022iss}.
At the opposite end of the non-Lorentzian spectrum lies Carrollian string theory, associated with the limit in which the speed of light is sent to zero~\cite{CardonaGomisPons2016}.
At the classical level, Carrollian strings have been constructed from several relativistic starting points, which lead to inequivalent Carrollian strings~\cite{CardonaGomisPons2016,Casalbuoni2024Carroll,Harksen2024Carroll}. 
Beyond the classical level, a Carrollian counterpart of nonrelativistic string theory has been identified as a decoupling limit of type II string theory~\cite{Blair2024}.
More recently, quantum aspects of Carrollian strings have also been explored, including studies of spectra, scattering amplitudes, and BRST quantization~\cite{Chen:2025gaz,Figueroa-OFarrill:2025njv}.
Carroll symmetry also enters string theory through the tensionless limit, in which the worldsheet rather than the target space becomes Carrollian~\cite{LST1991,ILST1994,Bagchi:2013bga,Bagchi:2015nca}.
Recent work has explored the Carroll--Weyl gauge structure of null strings and its interplay with Carrollian target geometries~\cite{SheikhJabbariYavartanooCW2026,SheikhJabbariYavartanooClassical2026,SheikhJabbariYavartanooTargets2026}.

Galilean and Carrollian structures arise from opposite limits, yet their relation becomes particularly transparent in two dimensions, where the two are exchanged by interchanging the temporal and spatial directions~\cite{Duval:2014uoa,Athanasiou2024}, and a Lie-algebraic duality survives in higher dimensions~\cite{FigueroaOFarrill2023}.
For strings, the Galilean and Carrollian worldsheet theories of the nonrelativistic string are exchanged by trading the longitudinal temporal and spatial vielbeine~\cite{Bidussi2023Longitudinal}. 
For branes, Galilean and Carrollian limits are related by interchanging longitudinal and transverse directions~\cite{Bergshoeff:2020xhv}, in particular when time is moved from the worldvolume to the transverse space~\cite{Bergshoeff:2026alice}. In double field theory, torsional Newton--Cartan and Carrollian geometries are non-Riemannian backgrounds of one $O(D,D)$-covariant framework~\cite{MorandPark2017,ChoPark2020,Blair:2020gng,LescanoOsten2024}, related by T-duality~\cite{Gallegos2021}.
These developments place the two limits within larger string-theoretic structures, including the web of decoupling limits~\cite{Blair2024,GomisYan2024}, as well as matrix-theory decoupling limits and their non-Lorentzian target-space and supergravity descriptions~\cite{Blair:2024aqz,Blair:2025prd,Maskalaniec:2026vlk}, while retaining them as distinct sectors or distinct corners
rather than identifying both as complementary quotients
of a single non-Lorentzian superalgebra.

In this Letter we realize both limits within a single non-Lorentzian structure.
Starting from the type II superalgebras, we retain both the spacetime momentum $P_M$ and the fundamental-string winding charge $W_M$~\cite{deAzcarraga1989,BergshoeffVanProeyen2000,Sakaguchi1999} and subject the two vector charges to a single non-Lorentzian contraction.
The resulting non-Lorentzian master algebra contains a Galilean and a Carrollian translation module as ideals, and the Galilean and magnetic Carroll superalgebras are obtained as complementary quotients.
The two structures therefore coexist within the same contracted algebra before either branch is selected.
The same organization extends to the worldsheet, where nonrelativistic Green--Schwarz superstrings were previously constructed in ordinary superspace~\cite{GomisKamimuraTownsend2004,GomisGomisKamimura2005,GomisTownsend2017} and as a non-Riemannian background of a doubled formulation~\cite{Park:2016sbw}.
The doubled Green--Schwarz action~\cite{Bandos2015} admits the non-Lorentzian contraction before a polarization is chosen, and complementary polarizations then place either the momentum or the winding sector in the canonical phase-space role.
Algebraic quotients and worldsheet polarizations thus provide two distinct but parallel ways of exposing the two branches within each type II realization.

Two features of the type II string make this possible.
The first is that the string carries two vector charges.
A single non-Lorentzian structure has to contain both a Galilean and a Carrollian translation module, and therefore two independent vector charges, while an ordinary supersymmetry algebra contains only the momentum.
The second is the form in which the winding charge enters the superalgebra.
Galilean supersymmetry is obtained from the Poincar\'e superalgebra by contraction, with the momentum as its translation module~\cite{Puzalowski1978,Bergshoeff:2014gja,GomisKamimuraTownsend2004,GomisTownsend2017}.
Magnetic Carroll supersymmetry is obtained differently: unlike its electric counterpart~\cite{Bergshoeff:2015wma}, it does not follow from a contraction of the Poincar\'e superalgebra.
It was first constructed directly, in two~\cite{Ravera:2022buz} and three~\cite{Zorba_2025} dimensions (see also~\cite{Zheng:2025rfe}), and its relativistic origin was traced only later to a twisted parent of Hull type, in which the vector charge enters the supercharge bilinear with opposite signs for the two supercharges~\cite{TwistedOrigin2026,Hull:1998vg}.
The same pattern appears without supersymmetry, where the Carrollian counterpart of the Bargmann particle is proposed to descend from Hull's type IIA$^*$ theory~\cite{Bergshoeff:2026alice}.
In a particle realization the momentum is the only vector available, and the twisted vector has to be supplied by hand.
In the type II superalgebras the winding charge enters with exactly this sign structure~\cite{Hull:1998vg,BergshoeffVanProeyen2000}.
The twisted vector that magnetic Carroll supersymmetry requires is therefore intrinsic to the string, and both non-Lorentzian supersymmetries descend from the ordinary type II algebra.

To make this explicit, we restrict the type II
superalgebras~\cite{BergshoeffVanProeyen2000,Sakaguchi1999} to their
momentum and fundamental-string winding sector, where the
anticommutators are
\begin{subequations}\label{eq:relativistic-TypeII}
\small
\begin{align}
\{Q,Q\}_{\rm IIA}
&=
\frac12\,\Gamma^M C^{-1} P_M
+\frac12\,\Gamma^M\Gamma_\star C^{-1} W_M ,
\label{eq:relativistic-TypeII-a}
\\
\{Q^I,Q^J\}_{\rm IIB}
&=
\frac12\,(\Gamma^M C^{-1})
\left(
\delta^{IJ}P_M
+(\tau_3)^{IJ}W_M
\right).
\label{eq:relativistic-TypeII-b}
\end{align}
\end{subequations}

Here $P_M$ is the spacetime momentum and $W_M$ the fundamental-string
winding charge. In IIB, $I,J=1,2$ label the two same-chirality
supercharges and $\tau_3$ is the third Pauli matrix acting on this
doublet. The full IIB vector sector is a triplet, of which we display
the momentum--F1 subsector. The twist is explicit in IIB. The
momentum enters with the identity on the doublet, while the winding
enters with $\tau_3=\mathrm{diag}(1,-1)$, a form of indefinite
signature. In IIA the same structure appears upon decomposing the
Majorana supercharge into its chiral parts,
$Q^\pm=\Lambda_\pm Q$ with $\Lambda_\pm=\tfrac12(1\pm\Gamma_\star)$,
in terms of which the vector sector of
\eqref{eq:relativistic-TypeII} takes the following schematic
form~\cite{BergshoeffVanProeyen2000}
\begin{align}
\{Q^\pm,Q^\pm\}&\sim \Gamma^M C^{-1}\,Z^\pm_M ,&
\{Q^+,Q^-\}&=0 \,,
\label{eq:chiral-split}
\end{align}
where $Z^\pm_M =P_M\mp W_M$. Each chiral supercharge thus sees the momentum with
the same sign and the winding with opposite signs, which is the
indefinite structure of IIB in a different guise. We carry out the
contraction first in IIA, where it singles out the time direction
alone and the quotients are the particle-like Galilean and magnetic
Carroll superalgebras, and then in IIB, where it singles out a
longitudinal two-plane.

Apart from the additional vector charge $W_M$, the bosonic parent
algebra has the standard Lorentzian structure. $P_M$ and $W_M$
transform as Lorentz vectors and commute with each other and with
the supercharges%
\footnote{We use mostly-plus signature, $\Gamma_\star=\Gamma^0\Gamma^1\cdots\Gamma^9$, and the antisymmetric charge-conjugation matrix, $\Gamma^{M\,T}=-C\Gamma^MC^{-1}$, for which $\Gamma^MC^{-1}$ is symmetric. In IIB, all displayed supercharge brackets are understood to be restricted to the  Majorana--Weyl doublet.}. 
We introduce the grading element $S$ and its projectors
\begin{equation}
\Pi_\pm=\tfrac12(1\pm S),\qquad
S=\Gamma_0\Gamma_\star,\qquad
Q_\pm=\Pi_\pm Q ,
\label{eq:IIAgrading}
\end{equation}
in terms of which the vector sector of \eqref{eq:relativistic-TypeII-a} reads
\begin{equation}
\begin{aligned}
\{Q_\pm,Q_\pm\}&=\tfrac12\Pi_\pm\left(-\Gamma_0C^{-1}P_0
   +\Gamma_i\Gamma_\star C^{-1}W_i\right),\\
\{Q_+,Q_-\}&=\tfrac12\Pi_+\left(\Gamma_iC^{-1}P_i
   -\Gamma_0\Gamma_\star C^{-1}W_0\right).
\end{aligned}
\label{eq:IIAgraded}
\end{equation}
The two charges have complementary support. $P_0$ and $W_i$ appear
in the diagonal brackets, $P_i$ and $W_0$ in the mixed one. We
therefore keep $P_0$, $W_i$ and $Q_+$ fixed, rescale $P_i$, $W_0$,
$M_{0i}$ and $Q_-$ by $\omega^{-1}$, and take $\omega\to\infty$.
Denoting the finite rescaled generators by the same symbols and
writing $G_i=\omega^{-1}M_{0i}$, the bracket $\{Q_-,Q_-\}$ is
suppressed by $\omega^{-2}$ and drops out, while the other two are
unchanged. The anticommutators of the contracted algebra are thus
\eqref{eq:IIAgraded} with $\{Q_-,Q_-\}=0$, and its non-vanishing
commutators are
\begin{align}
[G_i,P_0]&=P_i, & [G_i,W_j]&=\delta_{ij}W_0, \nonumber\\
[G_i,Q_+]&=-\tfrac12\Gamma_{0i}Q_- .
\end{align}

A closely related In\"on\"u--Wigner contraction appears in
Ref.~\cite{Blair:2024aqz} in the construction of the M0T
superalgebra associated with the D0-brane decoupling limit
underlying BFSS matrix theory. Its momentum--F1 subsector
reproduces our contracted IIA algebra, up to conventions.%
\footnote{In Sec.~2.3 of that reference, $N=\tfrac{\epsilon^2}{2}(\mathcal P^0+\mathcal A_0)$ is the central charge of the super-Bargmann algebra of the D0-brane sector. Setting $\mathcal A_0=\mathcal A_2^{\mu\nu}=0$ and discarding higher-rank charges gives $N,N^i\to0$ for finite $H,H^i$. Their $(H,P^i)$ and $(H^i,P)$ then correspond to our $(P_0,P_i)$ and $(W_i,W_0)$, respectively, up to signs and normalization, their F1 charge $\mathcal A_1^\mu$ to $W_M$, and their contraction parameter $\epsilon$ to $\omega^{-1}$. Identifying their $\gamma_\mu$ with our $\Gamma_\mu$, one has $\gamma_0\gamma_{11}=-\Gamma_0\Gamma_\star$, so their unscaled supercharge lies in the opposite eigenspace. The two labelings are related by $Q\mapsto\Gamma_\star Q$, an automorphism of the momentum--F1 parent superalgebra that leaves $P_M$ and $W_M$ unchanged.}

The two translation modules $\langle P_0,P_i\rangle$ and
$\langle W_0,W_i\rangle$ are invariant ideals of this algebra.
Quotienting by the $W$-module sets $W_M=0$ in
\eqref{eq:IIAgraded} and leaves the Galilean superalgebra, in
which the Hamiltonian $P_0$ sits in the diagonal bracket and the
boosts act as $[G_i,P_0]=P_i$ with $[G_i,P_j]=0$. Quotienting by the $P$-module sets $P_M=0$ and leaves
the magnetic Carroll superalgebra, in which the Hamiltonian $W_0$
sits only in the mixed bracket and the boosts act as
$[G_i,W_j]=\delta_{ij}W_0$. The latter is the ten-dimensional
counterpart of the three-dimensional algebra constructed
in~\cite{Zorba_2025}, with the winding charge as its translation
module. Both non-Lorentzian supersymmetries are thus present in the
same contracted algebra. This algebra is the classical algebra of
charges. Since $\{Q_-,Q_-\}=0$, a representation with self-adjoint
$Q_-$ on a positive-definite space would force $Q_-$ to act
trivially, together with $P_i$ and $W_0$. A nontrivial unitary
realization therefore requires a central extension by winding
contributions, as for the Galilean
superstring~\cite{GomisTownsend2017}, or a sector in which $Q_-$ is
pure gauge. We do not address this here.

The IIB realization follows the same pattern, with a grading that
singles out a longitudinal two-plane. The parity of the number of
directions is fixed by chirality. Complementary support requires
the grading element to anticommute with the matrix that multiplies
the winding. In IIA this matrix is $\Gamma_\star$, which
anticommutes with products of an odd number of gamma matrices, so
the grading singles out an odd number of directions, and the
smallest choice is time alone. In IIB it is $\tau_3$, so the grading
carries $\tau_1$ and, to act within the chiral doublet, an even
number of gamma matrices. For the minimal choice of two gamma
matrices, requiring the grading element to square to one selects
a Lorentzian two-plane. We split $M=(A,a)$ with $A=(0,z)$ and
$a=1,\ldots,8$, write the two supercharges as the doublet
$Q=(Q^1,Q^2)^T$, and introduce
\begin{equation}
\mathcal P_\pm=\tfrac12(1\pm S),\qquad
S=\Gamma_0\Gamma_z\tau_1,\qquad
Q_\pm=\mathcal P_\pm Q ,
\label{eq:IIBgrading}
\end{equation}
with $\tau_1$ the first Pauli matrix on the doublet index and
$\mathcal C=C\otimes\mathbf 1_2$. In this basis the vector sector of
\eqref{eq:relativistic-TypeII-b} reads
\begin{equation}
\begin{aligned}
\{Q_\pm,Q_\pm\}&=\tfrac12\mathcal P_\pm\left(\Gamma^A\mathcal C^{-1}P_A
   +\Gamma^a\mathcal C^{-1}\tau_3W_a\right),\\
\{Q_+,Q_-\}&=\tfrac12\mathcal P_+\left(\Gamma^a\mathcal C^{-1}P_a
   +\Gamma^A\mathcal C^{-1}\tau_3W_A\right).
\end{aligned}
\label{eq:IIBgraded}
\end{equation}
The complementary support is now $P_A$, $W_a$ in the diagonal
brackets and $P_a$, $W_A$ in the mixed one. We keep $P_A$, $W_a$ and
$Q_+$ fixed, rescale $P_a$, $W_A$, $M_{Aa}$ and $Q_-$ by
$\omega^{-1}$, and take $\omega\to\infty$. With
$G_{Aa}=\omega^{-1}M_{Aa}$, the bracket $\{Q_-,Q_-\}$ again drops
out. The anticommutators of the contracted algebra are
\eqref{eq:IIBgraded} with $\{Q_-,Q_-\}=0$, and its non-vanishing
commutators are
\begin{align}
[G_{Aa},P_B]&=-\eta_{AB}P_a, & [G_{Aa},W_b]&=\delta_{ab}W_A, \nonumber\\
[G_{Aa},Q_+]&=-\tfrac12\Gamma_{Aa}Q_- .
\end{align}
The $P$- and $W$-modules are again invariant ideals. Quotienting by
the $W$-module gives the string-Galilei superalgebra, with the
longitudinal momenta $P_A$ in the diagonal bracket and the boosts
acting as $[G_{Aa},P_B]=-\eta_{AB}P_a$. Quotienting by the
$P$-module gives its magnetic Carroll counterpart, with the
longitudinal winding $W_A$ only in the mixed bracket and
$[G_{Aa},W_b]=\delta_{ab}W_A$.

The two contractions are related by spacelike T-duality.
Compactifying the direction $z$, the T-duality map $T_z$ exchanges
momentum and winding along $z$ and leaves the other components
unchanged,
\begin{equation}
P_z^{\mathrm{IIA}}\longleftrightarrow W_z^{\mathrm{IIB}},\qquad
W_z^{\mathrm{IIA}}\longleftrightarrow P_z^{\mathrm{IIB}},
\label{eq:Tduality-charges}
\end{equation}
while on a spinor parameter $\epsilon$ it acts
as~\cite{Hassan2000,KulikRoiban2002}
\begin{equation}
T_z\,\epsilon=\begin{pmatrix} \Lambda_+\epsilon\\
-\Gamma_z\Lambda_-\epsilon\end{pmatrix}. 
\label{eq:Tduality-spinor}
\end{equation}
This flips the chirality of one component, producing the same-chirality IIB doublet. The corresponding transformation of the supercharges maps the IIA grading into the IIB one,
\begin{equation}
\Gamma_0\Gamma_\star\ \xrightarrow{\ T_z\ }\ \Gamma_0\Gamma_z\tau_1 .
\label{eq:Tduality-grading}
\end{equation}
Under \eqref{eq:Tduality-charges} the IIA scaling, which keeps
$P_0$, $W_a$ and $W_z$ fixed, becomes the IIB scaling, which keeps
$P_A=(P_0,P_z)$ and $W_a$ fixed. Together with
\eqref{eq:Tduality-grading}, this maps the IIA  contracted algebra
onto the IIB one on the generators the two theories share once $z$
is compact, namely the rotations $M_{ab}$, the boosts
$G_a^{\mathrm{IIA}}\leftrightarrow G_{0a}^{\mathrm{IIB}}$, the
translations and the supercharges, the Lorentz generators with a leg
along $z$ being broken by the compactification. The translation ideals of
either theory are mapped to ideals of the other with the compact
charge exchanged,
\begin{equation}
\begin{aligned}
\langle P_0,P_a,P_z\rangle_{\mathrm{IIA\,(IIB)}}
&\xrightarrow{\ T_z\ }\langle P_0,P_a,W_z\rangle_{\mathrm{IIB\,(IIA)}},\\
\langle W_0,W_a,W_z\rangle_{\mathrm{IIA\,(IIB)}}
&\xrightarrow{\ T_z\ }\langle W_0,W_a,P_z\rangle_{\mathrm{IIB\,(IIA)}} .
\end{aligned}
\label{eq:Tduality-ideals}
\end{equation}
In each theory the image differs from the ideal obtained by its own
contraction by the exchange of $P_z$ and $W_z$. Both compact charges
commute with the common rotations and boosts, so each image is again
an ideal of the common algebra, and the boost relations
$[G_{0a},P_0]=P_a$ and $[G_{0a},W_b]=\delta_{ab}W_0$ hold in it
unchanged. Spacelike T-duality therefore preserves the Galilean or
magnetic Carroll structure of a quotient in either direction and
exchanges only the assignment of the compact momentum and winding.

We now give the worldsheet description of the same contraction. Our
starting point is the doubled Green--Schwarz action
of~\cite{Bandos2015}, in which the ordinary and dual
coordinates, $X^M=X_L^M+X_R^M$ and $Y_M=\eta_{MN}(X_L^N-X_R^N)$,
enter on the same footing. We fix the auxiliary Pasti--Sorokin--Tonin (PST) scalar to
worldsheet time, $a=\tau$, which brings the action to the first-order
form used below. In IIA the two opposite-chirality Weyl spinors
combine into a Majorana spinor $\Theta$, and in IIB they form a
doublet of Majorana--Weyl spinors of the same chirality. With
$\bar\Theta=\Theta^T C$ in IIA and $\bar\Theta=\Theta^T\mathcal C$ in
IIB, we define
\begin{equation}
\begin{aligned}
A^M&=i\bar\Theta\Gamma^M d\Theta,
&
\widetilde A_M&=i\bar\Theta\Gamma_M\Sigma\,d\Theta,
\\
\Pi^M&=dX^M-A^M,
&
\widetilde\Pi_M&=dY_M-\widetilde A_M,
\end{aligned}
\label{gs:parent-forms}
\end{equation}
where $\Sigma=\Gamma_\star$ for IIA and $\Sigma=\tau_3$ for IIB. In this notation the relativistic action is
\begin{equation}
\begin{aligned}
S_{\mathrm{rel}}={}&\int d\tau\,d\sigma\,
\biggl[\frac{T_{\mathrm{rel}}}{2}
\left(\Pi_\tau^M\widetilde\Pi_{M\sigma}+\widetilde\Pi_{M\tau}\Pi_\sigma^M\right)
\\
& -T_{\mathrm{rel}}\,u\,\Pi_\sigma^M\widetilde\Pi_{M\sigma}
-\frac{e_{\mathrm{rel}}T_{\mathrm{rel}}^2}{2}
\left(\Pi_\sigma^2+\widetilde\Pi_\sigma^2\right)\biggr]\\
&+\frac{T_{\mathrm{rel}}}{2}\int
\left(dX^M\wedge\widetilde A_M+dY_M\wedge A^M\right).
\end{aligned}
\label{gs:parent}
\end{equation}
Here $T_{\mathrm{rel}}$ is the string tension, $u$ and
$e_{\mathrm{rel}}$ are worldsheet multipliers, the squares are
contracted with the Minkowski metric, and the last integral is the
Wess--Zumino term. The action retains the full fermionic dependence
and inherits supersymmetry and local $\kappa$ symmetry from the
doubled action~\cite{Bandos2015}. The conversion from the variables
used there, including the identification of the multipliers, is
given in the Supplemental Material.

To take the non-Lorentzian limit in both theories at once, we write
$(\lambda,r)=(0,i)$ for IIA and $(\lambda,r)=(A,a)$ for IIB, and
split the fermions into eigenspinors of the grading element,
$S\theta_\pm=\pm\theta_\pm$. We then set
\begin{equation}
\begin{aligned}
X^\lambda&=x^\lambda, & X^r&=\omega^{-1}x^r,\\
Y_\lambda&=\omega^{-1}y_\lambda, & Y_r&=y_r,\\
\Theta&=\theta_++\omega^{-1}\theta_-, & T_{\mathrm{rel}}&=\omega T,\\
e_{\mathrm{rel}}&=\omega^{-2}e, & &
\end{aligned}
\label{gs:scaling}
\end{equation}
with $x^0=t$ in both theories. The unscaled coordinates, $x^0$ and
$y_i$ in IIA and $x^A$ and $y_a$ in IIB, are conjugate to the charges
kept fixed in the algebra, $P_0,W_i$ and $P_A,W_a$. The finite
fermionic one-forms are
\begin{equation}
\begin{aligned}
\mathcal A^\lambda
&=i\bar\theta_+\Gamma^\lambda d\theta_+,
\\
\mathcal A^r
&=i\bar\theta_+\Gamma^r d\theta_-
+i\bar\theta_-\Gamma^r d\theta_+,
\\
\mathcal B_\lambda
&=i\bar\theta_+\Gamma_\lambda\Sigma\,d\theta_-
+i\bar\theta_-\Gamma_\lambda\Sigma\,d\theta_+,
\\
\mathcal B_r
&=i\bar\theta_+\Gamma_r\Sigma\,d\theta_+.
\end{aligned}
\label{gs:finite-fermion-forms}
\end{equation}
Defining
\begin{equation}
E^M=dx^M-\mathcal A^M,
\qquad
F_M=dy_M-\mathcal B_M,
\label{gs:finite-forms}
\end{equation}
and taking \(\omega\to\infty\), we obtain the non-Lorentzian action
\begin{equation}
\begin{aligned}
S_{\mathrm{NL}}={}&\int d\tau\,d\sigma\,
\biggl[
\frac T2
\left(E_\tau^M F_{M\sigma}+F_{M\tau}E_\sigma^M\right)
\\
&\qquad
-Tu\mathcal D-\frac{eT^2}{2}\mathcal H
\biggr]
+\int B_{\mathrm{WZ}},
\end{aligned}
\label{gs:finite-action}
\end{equation}
where $\mathcal H$ is $\mathcal H_{\mathrm{IIA}}$ or $\mathcal H_{\mathrm{IIB}}$, according to the theory,
\begin{equation}
\begin{aligned}
B_{\mathrm{WZ}}
&=\frac T2
\left(
dx^M\wedge\mathcal B_M
+dy_M\wedge\mathcal A^M
\right),
\\
\mathcal D&=E_\sigma^M F_{M\sigma},
\\
\mathcal H_{\mathrm{IIA}}
&=F_{i\sigma}F_{i\sigma}-(E_\sigma^0)^2,
\\
\mathcal H_{\mathrm{IIB}}
&=\eta_{AB}E_\sigma^A E_\sigma^B
+F_{a\sigma}F_{a\sigma}.
\end{aligned}
\label{gs:finite-constraints}
\end{equation}
The limit is finite without subtracting any divergent term. The
terms that would diverge are those in which an unscaled coordinate
multiplies the leading fermion bilinear of its dual,
$\bar\theta_+\Gamma^r d\theta_+$ or
$\bar\theta_+\Gamma_\lambda\Sigma\,d\theta_+$, and these bilinears
vanish identically as
$\Pi_+^TC=C\Pi_-$ and both $\Gamma^r$ and $\Gamma_\lambda\Sigma$
commute with the grading element. This is the complementary support of the algebra
acting on the worldsheet. All fermion terms are retained, including
those quartic in the fermions, and the multipliers impose
$\mathcal D=\mathcal H=0$. The finite action retains supersymmetry
and local $\kappa$ symmetry. On the constraint surface
$\mathcal D=\mathcal H=0$ and wherever $E^0_\sigma\neq0$, the
limiting $\kappa$ transformations act on the fermions with rank
sixteen, as in the parent theory. The transformations and the rank
computation are given in the Supplemental Material.

The Galilean and Carrollian structures appear upon choosing either
$x^M$ or $y_M$ as the canonical coordinates, which are the two
polarizations of the doubled action. Writing a prime for
$\partial_\sigma$, the corresponding momenta are
\begin{equation}
p_M=Ty'_M,
\qquad
\pi^M=T(x^M)',
\label{gs:canonical-momenta}
\end{equation}
and the bosonic parts of the Hamiltonian constraints, $\mathscr C=T^2\mathcal H$, read
\begin{equation}
\begin{aligned}
\mathscr C_X^{\mathrm{IIA}}
&=p_i^2-T^2t'^2,
\\
\mathscr C_Y^{\mathrm{IIA}}
&=T^2y_i'^2-(\pi^0)^2,
\\
\mathscr C_X^{\mathrm{IIB}}
&=p_a^2+T^2\eta_{AB}(x^A)'(x^B)',
\\
\mathscr C_Y^{\mathrm{IIB}}
&=\eta_{AB}\pi^A\pi^B+T^2y_a'^2.
\end{aligned}
\label{gs:polarizations}
\end{equation}
The two constraints of each theory are the same function of the doubled fields and differ only in which variables are coordinates and which are momenta. In the IIA $X$ description the spatial momenta are set
against the derivative of time, which is the constraint of the
Galilean superstring~\cite{GomisTownsend2017}, although the fermionic
scaling and the supersymmetry transformations differ here. In the $Y$
description the temporal momentum is set against the derivatives of
the spatial coordinates, which is the Carrollian form. The IIB
descriptions follow the same pattern, with the longitudinal
two-plane in place of the single time direction. The IIA and IIB
$Y$-constraints reproduce the particle-like and stringy Carroll
string constraints of~\cite{CardonaGomisPons2016}, respectively.
Since the longitudinal momenta $p_A$ do not enter $\mathscr C_X^{\mathrm{IIB}}$,
their equations set $\dot x^A=u\,x'^A$, as $\dot t=u\,t'$ does in
the Galilean superstring, of which this description is the T-dual
along $z$. At the bosonic level, both $X$ descriptions belong to
the non-vibrating class of~\cite{GomisTownsend2017}, rather than to
the vibrating nonrelativistic string theory of~\cite{GomisOoguri2001},
whose limit requires a critical Kalb--Ramond field. In either polarization the bosonic action is first order,
\begin{equation}
\begin{aligned}
S_X&=\int d\tau\,d\sigma
\left(p_M\dot x^M-u p_M x^{M\prime}
-\frac e2\mathscr C_X\right),\\
S_Y&=\int d\tau\,d\sigma
\left(\pi^M\dot y_M-u\pi^M y'_M
-\frac e2\mathscr C_Y\right).
\end{aligned}
\label{gs:polarized-action}
\end{equation}
The fermionic terms in both descriptions
are fixed by \eqref{gs:finite-action}. The passage from the symmetric
kinetic term of \eqref{gs:finite-action} to either canonical form
involves a total derivative, which must be kept in winding and
boundary sectors. The constraints therefore carry the bosonic
content of each description. Each polarization
retains both translation sectors, the one not in the canonical role
entering as the monodromy of the canonical coordinate:
$W_M=T\oint (x^M)'\,d\sigma$ in the $X$ description and
$P_M=T\oint y'_M\,d\sigma$ in the $Y$ description. In the Galilean
polarization $W_0=T\oint t'\,d\sigma$ is therefore nonzero only
when $t$ is periodically identified and the string winds in time, or
for open strings. This is the sector of the Galilean
superstring~\cite{GomisTownsend2017}, where the same quantity is the
topological central charge that bounds the momentum, and it is also
the regular branch $E^0_\sigma\neq0$ of the $\kappa$ symmetry, since
$t'$ cannot be nonzero everywhere on a closed string otherwise. In
the Carrollian polarization the Carroll Hamiltonian is the momentum
conjugate to $y_0$, which the constraint $\mathscr C_Y^{\mathrm{IIA}}$
sets to $|\pi^0|=T|y_i'|$. Wherever $y_i'\neq0$, the constraint
implies $\pi^0\neq0$, with no condition on $y_0$, and the local
$\kappa$ symmetry has rank sixteen on the constraint surface. An
algebraic quotient instead removes
one sector. The quotients and the polarizations thus expose the two
structures in different ways.

Spacelike T-duality also relates the complete finite
actions~\cite{Hassan2000,KulikRoiban2002}. The map $T_z$ of
\eqref{eq:Tduality-spinor}, applied to $\Theta$, takes the IIA
fermion bilinears into the IIB ones with the two charges along $z$
exchanged,
\begin{equation}
\begin{aligned}
\bar\Theta\Gamma^z\Gamma_\star d\Theta
&\ \xrightarrow{\ T_z\ }\ \bar\Theta\Gamma^z d\Theta ,\\
\bar\Theta\Gamma^z d\Theta
&\ \xrightarrow{\ T_z\ }\ \bar\Theta\Gamma^z\tau_3\, d\Theta ,
\end{aligned}
\label{gs:bilinear-map}
\end{equation}
and leaves the bilinears in the other directions unchanged. On the
finite fields the duality therefore acts as
\begin{equation}
\begin{aligned}
x_{\mathrm{IIB}}^z&=y_{\mathrm{IIA},z},
&
y_{\mathrm{IIB},z}&=x_{\mathrm{IIA}}^z,
\\
\theta_\pm^{\mathrm{IIB}}
&=T_z\theta_\pm^{\mathrm{IIA}},
\\
E_{\mathrm{IIB}}^z&=F_{\mathrm{IIA},z},
&
F_{\mathrm{IIB},z}&=E_{\mathrm{IIA}}^z,
\end{aligned}
\label{gs:duality}
\end{equation}
with the coordinates in the other directions unchanged. This map
preserves the first-order kinetic term and $\mathcal D$, takes
$\mathcal H_{\mathrm{IIA}}$ into $\mathcal H_{\mathrm{IIB}}$, and
exchanges the two compact contributions to the Wess--Zumino term. It
therefore relates the full local actions, including their fermion
terms and fermionic symmetries.

In canonical variables the exchange of the compact momentum and
winding reads
\begin{equation}
p_z^{\mathrm{IIA}}=T(x_{\mathrm{IIB}}^z)',
\qquad
p_z^{\mathrm{IIB}}=T(x_{\mathrm{IIA}}^z)',
\label{gs:canonical-duality}
\end{equation}
and the canonical Lagrangian densities agree up to a total
derivative,
\begin{equation}
T_z\!\left(\mathcal L^{\mathrm{IIA}}_{X,Y}\right)
=\mathcal L^{\mathrm{IIB}}_{X,Y}
\pm T\left[\partial_\tau\!\left(y_z\,x^{z\prime}\right)
-\partial_\sigma\!\left(y_z\,\dot x^z\right)\right]_{\mathrm{IIB}},
\label{gs:boundary}
\end{equation}
with the upper sign for the $X$ description and the lower sign for
the $Y$ description. The total derivative must be kept when
comparing boundary conditions and winding sectors, and the compact
radii of the two finite theories are related by
$r_{\mathrm{IIA}}\,r_{\mathrm{IIB}}=1/(2\pi T)$.

In this Letter we have shown that the Galilean and magnetic Carroll
supersymmetries of string theory descend from a single contraction
of the type II superalgebra, with the momentum and the
fundamental-string winding charge as the two translation modules,
and that the same contraction is finite on the doubled
Green--Schwarz action, where the two supersymmetries appear as its
two polarizations. The IIA and IIB Carrollian polarizations of
\eqref{gs:finite-action} are Carrollian superstrings in
Green--Schwarz form, with magnetic Carroll supersymmetry and local
$\kappa$ symmetry, whose bosonic Hamiltonian constraints reproduce
those of the particle-like and stringy Carroll strings
of~\cite{CardonaGomisPons2016}, respectively. Their spectra and
quantum consistency are the next things to study. Such studies would
supply the Carrollian corner with what nonrelativistic string theory
already has. At the relativistic level the IIA algebra with both
charges descends from the M-algebra, the winding charge arising as
the component $Z_{\mu\,11}$ of the membrane charge, that is, from a
membrane wrapped on the M-theory
circle~\cite{BergshoeffVanProeyen2000}. We expect the contraction to
lift to eleven dimensions, with the two ideals descending from the
momentum and the membrane charge. The momentum--F1 sector is also only one
two-vector subsector of the full type II algebras. The D-brane
charges supply further vector and higher-rank charges, and S-duality
rotates the IIB string structure $\tau_3$ into the D1 structure
$\tau_1$, so we expect a web of non-Lorentzian ideals labelled by brane charges rather than a single pair, related on the Galilean side to the matrix $p$-brane decoupling limits~\cite{Blair2024,GomisYan2024,Blair:2024aqz}. Carroll
geometry has also been placed in the bulk of candidate de Sitter and
flat-space holography~\cite{Blair:2025nno}, and a Carrollian
superstring with spacetime supersymmetry and local $\kappa$ symmetry
is the kind of worldsheet object such constructions require. These
results trace the
Galilean and magnetic Carroll supersymmetries of string theory to a
single contraction of the type II algebra, with the two vector
charges of the string as its two translation modules. The dualities
that relate the charges relate the supersymmetries, and the web of
non-Lorentzian supersymmetries should be as wide as the web of
dualities that generates it.

\begin{acknowledgments}
M.O. is supported in part by TUBITAK grant 125F471 and Istanbul
Technical University Research Fund under grant number TGA-2025-46976.
M.O. also acknowledges support from the TUBITAK Incentive Award. U.Z.
is supported by the Scientific and Technological Research Council of
T\"urkiye (T\"UB\.ITAK) under Grant Nos. 125F467 (ARDEB 1001) and
125F024 (ARDEB 3501). During this work, the authors used the large
language models Astra and Fable 5.1 to assist with symbolic
calculations, to carry out and check parts of the derivations under
the authors' direction, and to draft and edit parts of the manuscript.
All results were independently checked by the authors, who take full
responsibility for the content of the paper.
\end{acknowledgments}

\bibliographystyle{utphys}
\bibliography{Refs}

\clearpage
\onecolumngrid
\begin{center}
{\large\bfseries Supplemental Material}
\end{center}
\setcounter{equation}{0}
\renewcommand{\theequation}{S\arabic{equation}}
\setcounter{section}{0}

We use the conventions of the Letter. The signature is mostly plus,
$\Gamma_\star=\Gamma^0\Gamma^1\cdots\Gamma^9$, and the charge-conjugation
matrix is antisymmetric with $\Gamma_M^T=-C\Gamma_MC^{-1}$.
The grading element is $S=\Gamma_0\Gamma_\star$
in IIA and $S=\Gamma_0\Gamma_z\tau_1$ in IIB, with projectors
$\Pi_\pm=\tfrac12(1\pm S)$, and $\Sigma=\Gamma_\star$ in IIA and
$\Sigma=\tau_3$ in IIB.
In the IIB formulas below, $\Pi_\pm$ and $C$
stand for the doublet-space matrices
$\mathcal P_\pm$ and $\mathcal C$, respectively.
The matrix $\Sigma$ anticommutes with $S$, and
$S^TC=-CS$, so that $\Pi_\pm^TC=C\Pi_\mp$. For a matrix $G$
the bilinear $\Pi_s^TCG\Pi_t$ vanishes unless $t=s$ when $G$
anticommutes with $S$, and unless $t=-s$ when $G$ commutes with $S$.
Worldsheet derivatives are $\dot f=\partial_\tau f$ and
$f'=\partial_\sigma f$, and $[U\wedge V]_{\tau\sigma}=U_\tau
V_\sigma-U_\sigma V_\tau$.

\section{From the doubled action to the first-order form}

The action of~\cite{Bandos2015} is written for the two chiral sectors
separately. Each sector carries a bosonic coordinate, $X_L^M$ or
$X_R^M$, a sixteen-component Majorana--Weyl spinor, $\theta^1$ or
$\theta^2$, and the supersymmetric one-forms, written here in our
conventions for the bilinears,
\begin{equation}
L^M=dX_L^M-A_L^M,\qquad
R^M=dX_R^M-A_R^M,\qquad
A_L^M=i\bar\theta^1\Gamma^Md\theta^1,\qquad
A_R^M=i\bar\theta^2\Gamma^Md\theta^2.
\label{sm:sectors}
\end{equation}
The two spinors have opposite chirality in IIA and the same chirality
in IIB. The Lagrangian two-form of each sector is of
Pasti--Sorokin--Tonin type~\cite{Pasti:1996vs},
\begin{equation}
\mathcal L_L=T_{\rm rel}\bigl[L^M\wedge v\,i_v(L_M-{*L_M})
+dX_L^M\wedge A_{LM}\bigr],
\qquad
\mathcal L_R=-T_{\rm rel}\bigl[R^M\wedge v\,i_v(R_M+{*R_M})
+dX_R^M\wedge A_{RM}\bigr],
\label{sm:pst}
\end{equation}
where $v=d\xi^m\,\partial_ma/|\partial a\cdot g\cdot\partial a|^{1/2}$
is built from the auxiliary scalar $a(\xi)$ and the worldsheet metric
$g_{mn}$, $*$ is the worldsheet Hodge dual, and $T_{\rm rel}$ is the
tension, set to one in~\cite{Bandos2015}. The scalar $a$ is pure gauge.
We fix $a=\tau$. Then $v$ has only a $\tau$ component and
$i_v{*L_M}$ is proportional to $L_{M\sigma}$ alone. Writing the
worldsheet metric as $ds^2=-N^2d\tau^2+h\,(d\sigma+N^\sigma d\tau)^2$,
the two-forms reduce to
\begin{equation}
\begin{aligned}
\mathcal L_L&=T_{\rm rel}\bigl[L_\tau\!\cdot\!L_\sigma-v_+L_\sigma^2
+[dX_L\wedge A_L]_{\tau\sigma}\bigr]\,d\tau\wedge d\sigma,
\\
\mathcal L_R&=-T_{\rm rel}\bigl[R_\tau\!\cdot\!R_\sigma-v_-R_\sigma^2
+[dX_R\wedge A_R]_{\tau\sigma}\bigr]\,d\tau\wedge d\sigma,
\end{aligned}
\qquad
v_\pm=N^\sigma\pm\frac{N}{\sqrt h}.
\label{sm:FJ}
\end{equation}
The conformal factor of the metric drops out, and the null directions
of the metric are $d\sigma+v_\pm d\tau=0$. This is the
Floreanini--Jackiw form of a chiral boson~\cite{Floreanini:1987as},
here for the full supersymmetric one-forms.

The doubled variables of the Letter are $X^M=X_L^M+X_R^M$,
$Y_M=\eta_{MN}(X_L^N-X_R^N)$, and $\Theta=\theta^1+\theta^2$ in IIA or
$\Theta=(\theta^1,\theta^2)$ in IIB. In both cases the bilinears of
\eqref{gs:parent-forms} are $A^M=A_L^M+A_R^M$ and
$\widetilde A_M=\eta_{MN}(A_L^N-A_R^N)$, so that $\Pi^M=L^M+R^M$ and
$\widetilde\Pi_M=\eta_{MN}(L^N-R^N)$. Substituting into
$\mathcal L_L+\mathcal L_R$ gives \eqref{gs:parent} with
\begin{equation}
u=N^\sigma,\qquad e_{\rm rel}T_{\rm rel}=\frac{N}{\sqrt h}.
\label{sm:multipliers}
\end{equation}
The multiplier $u$ is the shift of the worldsheet metric and
$e_{\rm rel}T_{\rm rel}$ is its lapse density, the two combinations
that multiply the Virasoro constraints in the Hamiltonian form of the
Polyakov string. The symplectic term and the two constraint terms of
\eqref{gs:parent} come from the first two terms in each bracket of
\eqref{sm:FJ}, with $v_\pm=u\pm e_{\rm rel}T_{\rm rel}$, and the
Wess--Zumino term is
$\tfrac12T_{\rm rel}(dX^M\wedge\widetilde A_M+dY_M\wedge A^M)
=T_{\rm rel}(dX_L^M\wedge A_{LM}-dX_R^M\wedge A_{RM})$. 
The relative sign between the kinetic and Wess--Zumino terms is essential for
$\kappa$-symmetry.

\paragraph*{Rigid supersymmetry.}
 The rigid supersymmetry of the finite action follows from the
parent one with $\epsilon=\epsilon_++\omega^{-1}\epsilon_-$,
$S\epsilon_\pm=\pm\epsilon_\pm$, and reads
\begin{equation}
\begin{aligned}
\delta_\epsilon\theta_\pm&=\epsilon_\pm,
&
\delta_\epsilon x^\lambda&=i\bar\epsilon_+\Gamma^\lambda\theta_+,
&
\delta_\epsilon x^r&=i(\bar\epsilon_+\Gamma^r\theta_-+\bar\epsilon_-\Gamma^r\theta_+),
\\
&&
\delta_\epsilon y_\lambda&=i(\bar\epsilon_+\Gamma_\lambda\Sigma\theta_-+\bar\epsilon_-\Gamma_\lambda\Sigma\theta_+),
&
\delta_\epsilon y_r&=i\bar\epsilon_+\Gamma_r\Sigma\theta_+ .
\end{aligned}
\label{sm:rigid-susy}
\end{equation}

The one-forms $E^M$ and $F_M$ are invariant,
and the Wess--Zumino term changes by a
total derivative. The transformations close
on translations, with
$[\delta_1,\delta_2]x^\lambda
=-2i\bar\epsilon_{1+}\Gamma^\lambda\epsilon_{2+}$.
Two minus-sector transformations commute.

The identity $\Pi_s^TC=C\Pi_{-s}$ implies that
$\epsilon_s$ pairs with the oppositely graded
supercharge. Writing $Q^{\rm rel}$ for the
relativistic supercharge, the finite worldsheet
charges are $q_-=\Pi_-Q^{\rm rel}$ and
$q_+=\omega^{-1}\Pi_+Q^{\rm rel}$, with
$\{q_+,q_+\}=0$ in the limit.
The redefinitions
$\widehat Q_\pm=\Gamma_\star q_\mp$ in IIA and
$\widehat Q_\pm=\tau_3 q_\mp$ in IIB reproduce
the contracted superalgebras of the Letter,
including the boost brackets, while leaving
all bosonic generators unchanged.

\section{Kappa symmetry and its rank}

\paragraph*{Parent theory.}
In a sector of \eqref{sm:FJ}, write $\vartheta$ for its spinor,
$K^M$ and $V^M$ for the $\sigma$ and $\tau$ components of its
one-form, and $v$ for its velocity. Under a variation $\delta\vartheta$
with $\delta X^M=i\bar\vartheta\Gamma^M\delta\vartheta$, the sector
Lagrangian changes, up to a total derivative, by
\begin{equation}
\delta\mathcal L=\pm T_{\rm rel}\bigl[-4i\,\delta\bar\vartheta\,
\Gamma_MK^M(\dot\vartheta-v\vartheta')-\delta v\,K^2\bigr].
\label{sm:parent-var}
\end{equation}
It vanishes for
\begin{equation}
\delta\vartheta=\Gamma_MK^M\kappa,\qquad
\delta v=4i\bar\kappa(\dot\vartheta-v\vartheta'),
\label{sm:parent-kappa}
\end{equation}
since $\delta\bar\vartheta=-\bar\kappa\Gamma_MK^M$ and
$(\Gamma_MK^M)^2=K^2$. The parameter has the chirality opposite to
$\vartheta$. In the doubled variables the two sectors combine into
\begin{equation}
\begin{aligned}
\delta_\kappa\Theta&=\tfrac12\bigl(\Gamma^M\Pi_{M\sigma}
+\Sigma\Gamma^M\widetilde\Pi_{M\sigma}\bigr)\kappa,
\qquad
\delta_\kappa X^M=i\bar\Theta\Gamma^M\delta_\kappa\Theta,
\qquad
\delta_\kappa Y_M=i\bar\Theta\Gamma_M\Sigma\,\delta_\kappa\Theta,
\\
\delta_\kappa u&=2i\bar\kappa\bigl(D_u\Theta
-e_{\rm rel}T_{\rm rel}\Sigma\Theta'\bigr),
\qquad
\delta_\kappa e_{\rm rel}=\frac{2i}{T_{\rm rel}}\,\bar\kappa\bigl(\Sigma D_u\Theta
-e_{\rm rel}T_{\rm rel}\Theta'\bigr),
\qquad
D_u=\partial_\tau-u\,\partial_\sigma,
\end{aligned}
\label{sm:doubled-kappa}
\end{equation}
where $\kappa$ is a Majorana spinor in IIA and a doublet of negative
chirality in IIB.

\paragraph*{Limit.}
The parameter scales oppositely to the coordinate spinor,
$\kappa=\omega^{-1}k_++k_-$ with $Sk_\pm=\pm k_\pm$. Inserting
\eqref{gs:scaling} into \eqref{sm:doubled-kappa}, the matrix acting on
$\kappa$ becomes $K_0+\omega^{-1}K_1+O(\omega^{-2})$ with
\begin{equation}
K_0=\tfrac12\bigl(\Gamma_\lambda E^\lambda_\sigma
+\Sigma\Gamma^rF_{r\sigma}\bigr),
\qquad
K_1=\tfrac12\bigl(\Gamma_rE^r_\sigma
+\Sigma\Gamma^\lambda F_{\lambda\sigma}\bigr).
\label{sm:K0K1}
\end{equation}
$K_0$ anticommutes with $S$ and $K_1$ commutes with it. Projecting,
and using $\theta_-=\omega\Pi_-\Theta$, the finite transformations are
\begin{equation}
\begin{aligned}
\delta_\kappa\theta_+&=K_0k_-,
\qquad
\delta_\kappa\theta_-=K_0k_++K_1k_-,
\qquad
\delta_\kappa x^M=\mathcal A^M\big|_{d\theta\to\delta_\kappa\theta},
\qquad
\delta_\kappa y_M=\mathcal B_M\big|_{d\theta\to\delta_\kappa\theta},
\\
\delta_\kappa u&=2i\,\bar k_-D_u\theta_+,
\qquad
\delta_\kappa e=\frac{2i}{T}\bigl[\bar k_+\Sigma D_u\theta_+
+\bar k_-\Sigma D_u\theta_--eT\,\bar k_-\theta_+'\bigr],
\end{aligned}
\label{sm:finite-kappa}
\end{equation}
where $\mathcal A^M$ and $\mathcal B_M$ are the bilinears of \eqref{gs:finite-fermion-forms}
with $d\theta_\pm$ replaced by $\delta_\kappa\theta_\pm$. The
variation of the rescaled parent Lagrangian is a total derivative at
every $\omega$ and has a finite expansion in $\omega^{-1}$, so its
leading coefficient is a total derivative. 
Hence the finite action \eqref{gs:finite-action} is invariant under \eqref{sm:finite-kappa}.

\paragraph*{Rank.}
We give the IIA computation. With $\ell=E^0_\sigma$, $v_i=F_{i\sigma}$,
$w_i=E^i_\sigma$ and $r=F_{0\sigma}$,
\begin{equation}
K_0=\tfrac12(\ell\Gamma_0-v_i\Gamma^i\Gamma_\star),
\qquad
K_1=\tfrac12(w_i\Gamma^i+rS),
\qquad
\mathcal H=v^2-\ell^2,
\qquad
\mathcal D=\ell r+v\cdot w.
\label{sm:IIAkappa}
\end{equation}
Let $V_\pm$ be the sixteen-dimensional eigenspaces of $S$. On
$\mathcal H=0$ with $\ell\neq0$, the matrix $J=\ell^{-1}v_i\Gamma^i$
squares to one and commutes with $S$. Rotating $v_i$ to a single axis,
a transverse gamma matrix preserves $V_s$ and anticommutes with $J$,
so each eigenvalue of $J$ has multiplicity eight on each $V_s$. With
$R_\pm=\tfrac12(1\pm J)$, one finds $K_0=\ell\Gamma_0R_{-s}$ on $V_s$.
Hence $K_0$ has rank eight on each $V_s$, and the first output of
\eqref{sm:finite-kappa} has rank eight. In the second output,
$K_0k_+$ spans $R_+V_-$. For $\zeta$ in the kernel $R_-V_-$ of $K_0$
on $V_-$, a direct Clifford multiplication gives
\begin{equation}
JK_1\zeta=K_1\zeta+\ell^{-1}\mathcal D\,\zeta.
\label{sm:JK1}
\end{equation}
On $\mathcal D=0$ this places $K_1\zeta$ in $R_+V_-$, which is already
spanned by $K_0k_+$. The kernel of the first output therefore adds no
new direction to the second, and the rank of
$(k_+,k_-)\mapsto(\delta_\kappa\theta_+,\delta_\kappa\theta_-)$ is
eight plus eight, which is sixteen. Both constraints are needed. On
$\mathcal H=0$ alone the rank is twenty-four at generic points. The
argument fixes the ranks on the bosonic body of the fields. Over the
Grassmann algebra $R_\pm$ remain idempotents and $\ell$ remains
invertible, so the ranks of the full even expressions are those of
their bodies. The condition $E^0_\sigma\neq0$ is the regular branch
quoted in the Letter.

The IIB transformations follow from \eqref{sm:K0K1} with
$\Sigma=\tau_3$, or from the IIA ones by T-duality. The $\kappa$
parameter is mapped by
$T^\kappa_z\,k=(\Lambda_-k,\,\Gamma_z\Lambda_+k)$, with the chiralities
and the sign opposite to the map of \eqref{eq:Tduality-spinor}, and with the dictionary
of \eqref{gs:duality} one checks $T_zK_0=K_0^{\rm IIB}\,T^\kappa_z$ and
$T_zK_1=K_1^{\rm IIB}\,T^\kappa_z$. The
rank sixteen is transported, and on the body of
$\mathcal H_{\rm IIB}=0$ a nonzero $(E^0_\sigma,E^z_\sigma,F_{a\sigma})$
has $E^0_\sigma\neq0$, which is the same regularity condition. The
rank refers to the fermionic coordinates. Parameters in its kernel can
still act on $u$ and $e$ by terms proportional to the fermionic field
equations, as in the parent theory.

\end{document}